\documentclass{article}
\usepackage{listings}
     \PassOptionsToPackage{numbers, compress}{natbib}

    \usepackage[preprint]{neurips_2023}

\usepackage[utf8]{inputenc} 
\usepackage[T1]{fontenc}    
\usepackage{hyperref}       
\usepackage{url}            
\usepackage{booktabs}       
\usepackage{amsfonts}       
\usepackage{nicefrac}       
\usepackage{microtype}      
\usepackage{xcolor}         
\usepackage{graphicx}
\usepackage{algorithm} 
\usepackage{algpseudocode} 

\title{Multimodal Gen-AI for \\Fundamental Investment Research}

\author{
  Lezhi Li$^*$ \\
  \texttt{lezhi@stanford.edu} \\
  \And
  Ting-Yu Chang$^*$ \\
  \texttt{tingyuc@stanford.edu} \\
  \AND
  Hai Wang$^*$ \\
  \texttt{haiw@stanford.edu} \\
}

\begin{document}

\maketitle
\def\thefootnote{*}\footnotetext{These authors contributed equally to this work.}

\begin{abstract}
  This report outlines a transformative initiative in the financial investment industry, where the conventional decision-making process, laden with labor-intensive tasks such as sifting through voluminous documents, is being reimagined. Leveraging language models, our experiments aim to automate information summarization and investment idea generation. We seek to evaluate the effectiveness of fine-tuning methods on a base model (Llama2) to achieve specific application-level goals, including providing insights into the impact of events on companies and sectors, understanding market condition relationships, generating investor-aligned investment ideas, and formatting results with stock recommendations and detailed explanations. Through state-of-the-art generative modeling techniques, the ultimate objective is to develop an AI agent prototype, liberating human investors from repetitive tasks and allowing a focus on high-level strategic thinking. The project encompasses a diverse corpus dataset, including research reports, investment memos, market news, and extensive time-series market data. We conducted three experiments applying unsupervised and supervised LoRA fine-tuning on the llama2\_7b\_hf\_chat as the base model, as well as instruction fine-tuning on the GPT3.5 model. Statistical and human evaluations both show that the fine-tuned versions perform better in solving text modeling, summarization, reasoning, and finance domain questions, demonstrating a pivotal step towards enhancing decision-making processes in the financial domain. Code implementation for the project can be found on GitHub: \url{https://github.com/Firenze11/finance\_lm}.
\end{abstract}

\section{Introduction}

In the financial investment industry, the process of decision-making is often cumbersome and resource-intensive, owing to the need to sift through an extensive array of documents. Investment analysts oftentimes find themselves inundated with company filings, research reports, and a plethora of market data signals. While these pieces of information are invaluable for informed decision-making, the labor involved in analyzing them is exceedingly demanding.

This report introduces a few experiments aimed at revolutionizing this process by leveraging language models. Our objective is to develop a system that automates the tasks of information summarization and investment idea generation. In particular, we would like to evaluate whether LLM fine-tuning methods that can help achieve the following application-level goals, and understand which fine-tuning strategies are most effective in achieving these goals:
\begin{enumerate}
    \item Given an event (which can come from a user’s question or be extracted from the latest news) such as oil price increase, answer which companies and sectors are impacted, and thereby generating investment ideas.
    \item Have reasonable “knowledge” about the relationships between various market conditions (valuation, emotions, liquidity, fundamentals etc) and the changes in various equity prices.
    \item Output investment ideas that are in line with the investors’ preferences and their investment thesis.
    \item Format the results in terms of: stock recommendation, top 3 reasons, and detailed explanation.
\end{enumerate}
By employing state-of-the-art generative modeling techniques, the end goal of this project aspires to produce a prototype of an AI agent for human investors, liberating them from low-value, repetitive analytical tasks, enabling them to allocate more time and resources on high-level strategic thinking.

\section{Related works}
\label{gen_inst}

The recent advancements in large language models (LLMs) form the cornerstone of our project. Our team, including a member with extensive financial industry experience, has witnessed a significant shift towards big data and traditional machine learning in investment research and trading. For instance, \citet{JP} at JP Morgan have been at the forefront of utilizing big data for research tailored to quantitative-focused investment clients for over a decade. The emergence and success of LLMs present an exciting opportunity to revolutionize the investment process by integrating advanced language-processing capabilities with existing data-driven approaches.

Our research encompasses the latest developments in this field, drawing from both academic sources and industry practices. We have analyzed academic studies that employ LLMs for time series forecasting. Notably,
\citet{gruver2023large} have explored the use of LLMs as zero-shot time series forecasters. Another noteworthy study from Taiwan conducted by \citet{chang2023llm4ts} demonstrates the enhancement of time-series forecasting through the fine-tuning of pre-trained LLMs.  \citet{yu2023temporal}, from an Amazon research group present a novel study on harnessing Large Language Models’ (LLMs) for explainable weekly and monthly financial forecasting. The experiments involve utilizing historical stock price data, company metadata, and economic/financial news to demonstrate the efficacy of large language models (LLMs)  in cross-sequence reasoning and inference. The results indicate that GPT-4 and fine-tuned Open LLaMA outperform traditional statistical/econometric models like ARMA-GARCH and gradient-boosting tree models, showcasing their ability to make informed decisions by integrating information from textual news and price time series.

In the industry sector, Bloomberg, the leading financial data and news provider, has contributed to this field with its BloombergGPT paper \citet{Bloomberg}, which investigates the integration of LLMs with financial data. Similarly, tech teams at Amazon have conducted intriguing experiments, leveraging LLMs for time-series forecasting using various sources, including public news and financial data.
 
Building on these findings, our project aims to go a step further. We are exploring the use of institutional research reports, typically exclusive to professional investors, to fine-tune LLMs. This approach seeks to develop novel methods for applying AI in finance, extending beyond the analysis of publicly available news.

\section{Problem statement}
\label{headings}

\subsection{Dataset}

We have amassed an extensive corpus dataset that includes research reports from leading global banks, investment memos, market news, and a substantial amount of time-series market data.

Project Exploration on language based investment data:
Our project is picking text data from a broad range of investment-related corpora. First, we are using top brokers’ research reports which can be further categorized from top-down research perspective: high-level macro \& market reports, middle-level sector research reports, and bottom-level equity research reports. In addition, we are integrating text data such as Investment Principles papers from global investment firms, Market Wraps received from research institutions, and widely available public news. The purpose and value of each category are detailed below:
\begin{enumerate}
    \item Macro \& Market Reports: These reports provide the LLM models many observations of macroeconomic and market trends, capturing human reasoning about the market dynamics. The objective is to train the LLM to understand the reasoning behind key macroeconomic factors in financial markets, such as inflation, commodities, and monetary policy etc. 
    \item Sector Reports: This part mainly includes professional research analysts work, which provides investors with very detailed specific industries’ overview, supply chains, sector growth, major players, and the regulatory environment.
    \item Equity Research Reports: These reports conduct thorough analyses of individual companies, focusing on the individual company’s business model, profitability, management team, valuation, and risks. After training, the LLM is anticipated to respond to fundamental equity-related questions.
    \item Investment Principles: These articles focused on the core philosophies, strategies, and processes guiding investment decisions. They are instrumental in synthesizing information from market, sector, and equity research, thereby reflecting the decision-making frameworks utilized by investors.
    \item Market Wraps: Offering professional updates on current market events, these can be regular emails or internal memo that provide expert viewpoints and address specific market dynamics. They play a crucial role in helping the LLM understand a variety of market dynamics and the perspectives of professional investors.
    \item News: This category includes updates on recent events and their varied impacts on markets, sectors, or equities. They are abundant sources of information, effectively representing the relationships and extensive knowledge for LLM to learn the details of the market.
\end{enumerate}
By explicitly training LLM models with different categories of text dataset, we expect the LLM models to learn different knowledge and have the capacity of Thoughts of Chain for investment purposes. 

Currently, we are using open-source tools to scrap a few Goldman Sachs (GS)  research portal pages, and convert PDF research reports into plain text for unsupervised fine-tuning. For supervised fine-tuning, we combine historical stock market news headlines data from kaggle \url{https://www.kaggle.com/datasets/miguelaenlle/massive-stock-news-analysis-db-for-nlpbacktests} with the daily stock returns data from the yfinance API \url{https://pypi.org/project/yfinance/}.

To fine-tune the GPT-3.5-turbo model, we compile a Q\&A dataset, primarily sourced from responses provided by portfolio managers and equity research analysts. This dataset encompasses three distinct categories of instructions: 1) Investment Philosophy (Approximately 10+ Instructions): This category delves into the investment philosophy adopted by a prominent value investing hedge fund. 2) Investment Methodologies part includes instructions on equity research processes and various other methodologies utilized in investment decision-making. 3)Facts and Knowledge Pertaining to Investments part consists of instructions aimed at providing factual information and knowledge relevant to investments. It encompasses sector-specific investment logic, details about specific companies within various sectors, and their respective names.

\section{Technical approach}
In the initial project stage, we carried out two experiments to evaluate the effect of model fine-tuning as compared to baseline, in the context of financial investment recommendations. 
\begin{enumerate}
    \item Unsupervised model fine-tuning: fine-tuning a base model (Llama2) using a pure text data corpus, on the task of next-token-prediction.
    \item Supervised model fine-tuning: fine-tuning the same base model using “text, stock price” pairs, using news headlines as text and the corresponding stock price change as labels.
    \item Instruction fine-tuning: fine-tuning GPT 3.5 turbo with the public API using a set of Question\&Answer instruction.
\end{enumerate}

In the first two scenarios, we used llama2\_7b\_hf\_chat as the base model, and LoRA for parameter-efficient fine-tuning framework. We used Hugging Face as the modeling library.

Our first step involves preparing text transformed directly from GS reports to train an Llama model. This is to get an initial understanding of the fine-tuning effect and potential improvements. We are also experimenting with time-series-based supervised training including stock price data etc. At this stage, we expect to produce enhanced responses to investment-related prompts. But there should be a noticeable gap between the model generated answer and that of an human analyst. 

Our next steps involve exploring and utilize other advanced models like GPT 3.5 Turbo for fine-tuning and so that we can comparing the performance of different models. Finally we refine our data selection process, potentially being focused on some specific sectors or training models on a particular investment strategy. We aim to enhance the relevant investment principles and rules, with the objective of evolving the LLM into a specialized equity analyst or portfolio manager. 

\subsection{Base model}
\begin{enumerate}
    \item Llama2: The model we chose for the initial experiments is llama2\_7b\_hf\_chat, a 7B parameter open source language model released by Meta. Unlike the plain Llama2, this version of Llama2 we chose is already fine-tuned by human feedback (“hf”) as well as instruction-fine-tuned to do well in a question-answering chat-bot like setting (“chat”). The reason we chose this model is not only to take advantage of its polished, human-friendly output basis, but also to evaluate how much more we can do to improve an already high baseline. During training, the model is loaded in 4-8 bit precision rather than the full 32 bit, for better training speed and lower cost.
    \item GPT3.5: GPT-3.5 models can understand and generate natural language or code. We chose the most capable and cost effective model in the GPT-3.5 family, gpt-3.5-turbo, which has been optimized for chat using the Chat Completions API but works well for traditional completions tasks as well. Fine-tuning lets us get more out of the models available through the API by providing higher quality results than prompting for financial questions, ability to train on more examples than can fit in a prompt and lower latency requests.
\end{enumerate}

\subsection{LoRA}
\begin{figure}[tb]
  \centering
  \includegraphics[width=15cm]{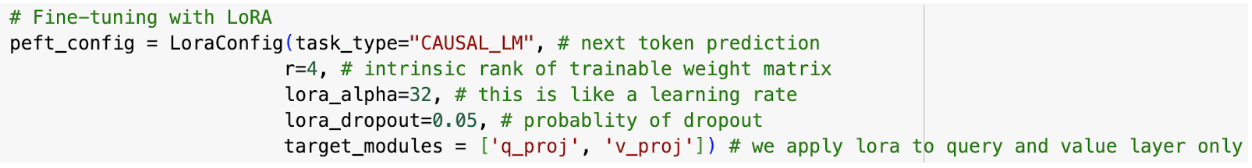}
  \caption{Detailed settings of LoRA.}
  \label{fig:LoRA}
\end{figure}

Low rank adaptation is proposed in \cite{hu2021lora} as an efficient way to fine-tune foundation models. LoRA works by adding auxiliary low-rank weight matrices next to the full-rank ones from the original foundation model, and only training the low-rank weights during fine-tuning. The activation in the forward pass is calculated by summing the output from these two weights. Formally:
$$ h = W_0 x + \Delta W x = W_0 x + BAx $$ 
This architecture introduces the benefit of faster training speed, lower cost, as well as protection over the base model from being overwritten unfavorably.
We used the LoRA implementation from the Hugging Face PEFT library, chose rank = 4 and modified the query and value matrices from the transformer layers only. See figure \ref{fig:LoRA} below for other detailed settings.

\subsection{Unsupervised fine-tuning}
\begin{figure}[tb]
  \centering
  \includegraphics[width=14cm]{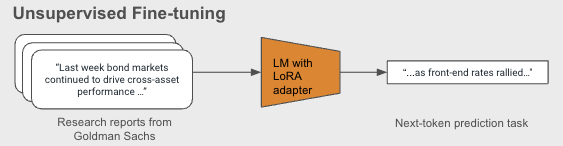}
  \caption{Unsupervised fine-tuning of Llama2 using LoRA}
  \label{fig:unsupervised_finetune}
\end{figure}

For the unsupervised training experiment, our goal is to feed the language model financial domain knowledge articles, so that the model picks up both the tone and the internal logic from domain experts. Figure \ref{fig:unsupervised_finetune} illustrates the unsupervised fine-tune process.

We trained with ~600 market analysis articles scraped from Goldman Sachs market analysis site. These articles are collected over the course of 2 days and cover topics ranging from individual stock analysis, to market segments, regions, and macro economics. Please see Appendix for an example text. Once the text was collected, we used it to train a next-token prediction model. The following pseudo code shows the detailed data manipulation as well as training process:
\begin{algorithm}
	\caption{Unsupervised fine-tuning training process} 
	\begin{algorithmic}[1]
     \State Load all articles from the scraped data directory;
     \State Concatenate all articles together;
     \State Tokenize all words;
     \State Chunk the concatenated corpus into chunks of 512 tokens each;
		\For {each chunk:}
			\For {each word:}
				\State Predict the next word given the context of all previous words;
            \EndFor
			\State Collect prediction loss across of the whole chunk;
		\EndFor
     \State Report training loss at the end of each epoch;
	\end{algorithmic} 
\end{algorithm}

We collected the training loss at the end of each epoch and compared the generation results with the baseline model.

\emph{Model loss:} due to time constraint, the training loss at the end of one epoch is around 2.29; eval loss was not compared.

\emph{Generation result:} we attached the text generation results from the fine-tuned vs baseline model in the appendix. Based on human evaluation, the fine-tuned result is much better than the baseline, and the analysis/reasoning in general does not have factual errors. This is in line with (and exceeds) what we expected.

\subsection{Supervised fine-tuning}

\begin{figure}[tb]
  \centering
  \includegraphics[width=14cm]{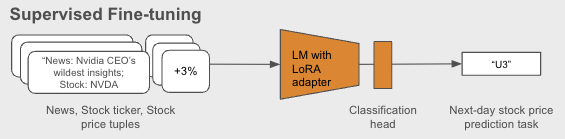}
  \caption{Supervised fine-tuning of Llama2 using LoRA, where the text comes from news headlines and targets are labels indicating stock price changes.}
  \label{fig:supervised_finetune}
\end{figure}

For the other experiment, we used news headlines to predict the corresponding stock price changes. Our news headline data is associated with the corresponding timestamp and the stock ticker mentioned in the news as metadata, and we used this information to query yfinance API to get the stock price for the ticker at the specific timestamp. 
We formulated the prediction task as a text classification task, with each headline text predicting stock price “up 3 percent” (U3), “down 1 percent” (D1) etc in the next day. We have a total of 12 labels ranging from “down 5+ percent” (D5+) to “up 5+ percent” (U5+). We used MSE loss for the classification error, with D5+ corresponding to -6, D5 corresponding to -5, etc. Figure \ref{fig:supervised_finetune} illustrates the training process.

\emph{Accuracy:} due to time constraint for data processing, we only use 1000 stocks tickers for now, the accuracy is around 1.3 which is higher than 1 with 100 stocks.

\emph{Generation result:} we use the recommended stocks provided by unsupervised learning to predict the stocks returns, however, the predictions are all "down 5+ percent" (D5+), which is the opposite as commonly expected. At this stage, we were experimenting with the text and time series relationship. Next we will put more efforts to uncover more robust relationships between various types of text and different forms of time series data.

\subsection{Instruction fine-tuning}
We conducted experiments on the instruction fine-tuning of the GPT-3.5 Turbo model with the goal of emulating the analytical capabilities of a human investment analyst. The model was tasked with conducting investment research and assessing investment opportunities, much like a portfolio manager. To facilitate this, we incorporated specific investment philosophies and equity research methodologies into a Q\&A format, subsequently converting these into a JSONL file for the model's training and application.

\begin{figure}[b]
  \centering
  \includegraphics[width=14cm]{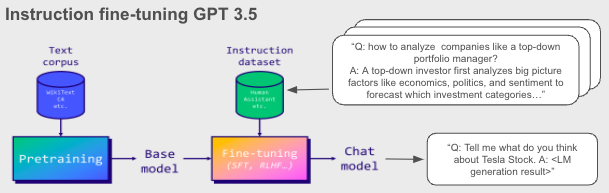}
  \caption{Instruction fine-tuning of GPT3.5 using question and answer pairs intended to impart investment principles.}
  \label{fig:instruction_finetune}
\end{figure}

During fine-tuning, we characterized the system as an 'investment research analyst chatbot that provides fundamental analysis of macro, market, sector, and equity.' This characterization aimed to better align the GPT model with its intended role. Figure \ref{fig:instruction_finetune} illustrates the training process.

We fine-tuned the model with just over 30 manual instructions, which resulted in noticeable changes in the model's responses. Through experimentation with various instruction combinations, we empirically discovered that a well-balanced and diversified set of instructions yields better results. The fine-tuned model generates responses that more closely adhere to investment logic and accurately reflect the reasoning processes of investment analysis.

\section{Results}

\subsection{Evaluation procedures}

With the final goal of helping investment managers making decisions in mind, we conducted three different types of model evaluations, each evaluating the performance at different abstraction levels:
\begin{enumerate}
    \item Evaluation of test set probability: We evaluated the next-token generation accuracy using a held-out test dataset, which is acquired using the same method as the training set. We think of this as an apple-to-apple comparison between the base model and fine-tuned model, and an evaluation of the fine-tuning procedure itself.
    \item Generalizability evaluation using publicly available datasets with ground-truth: we ran the models on some public benchmarking datasets, to evaluate the models capability on general language model tasks such as summarization and reasoning. The benefit of this step is that the ground truths in these datasets provide an objective measurement, and at the same time reflect how the model might perform in real world applications that they are intended for: summarizing information from investment reports and providing logical suggestions. 
    \item Human evaluation: we invited a group of investment professionals to rate the helpfulness of generated results. This evaluation is closest to the practical application of these models since it directly addresses investment professionals’ usage scenario.  It also compensates for the lack of finance-domain contents in the public benchmarking dataset corpus used above.
\end{enumerate}
For these three evaluations, we applied test set probability eval and generalizability eval on baseline Llama2 and the unsupervised fine-tuned version of Llama2. We also conducted human evaluation on baseline Llama2, unsupervised fine-tuned Llama2, gpt3.5, and instruction fine-tuned gpt3.5. Due to API limitations, we weren’t able to run test set probability evals on GPT models. 

\subsection{Evaluation of test set probability}
\begin{figure}[ht]
  \centering
  \includegraphics[width=10cm]{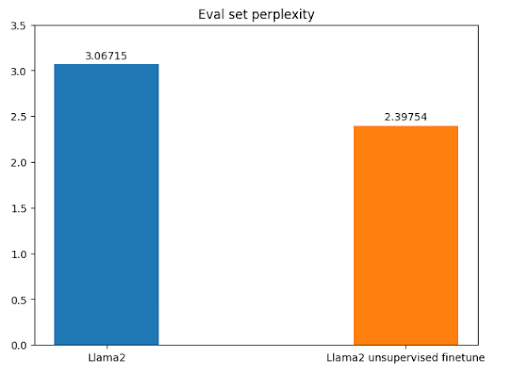}
  \caption{Comparison of eval set loss for the baseline and fine-tuned Llama2 models.}
  \label{fig:Loss}
\end{figure}

We first computed the eval loss on the held-out test set using the base Llama2 and the fine-tuned Llama2 models. The result is shown in Figure \ref{fig:Loss}. The test dataset is acquired in the same way as the training set – by scraping the Goldman Sachs investment research site, but over a different time period to ensure that the articles are disjoint from the training set articles. We can expect the test set distribution is consistent with the training set. A total of 244 articles were included for testing. 

The evaluation metric is perplexity, defined as
$$Loss = exp( -\frac{1}{t} \sum_{i = 0}^{t} log  p_{\theta}(x_i | x_{<i}))$$

where $p_{\theta}(x_i | x_{<i})$ is the probability of token $i$ given all previous contexts, computed by comparing model prediction logits with ground-truth next tokens in the original articles. We can see that the eval loss for the test dataset is lower than the training set, suggesting effective modeling of articles of similar genre / topics.

\subsection{Generalizability evaluation}
In addition to evaluation on the held-out test set, we chose two LLM benchmarking datasets with evaluation metrics that are aligned with the goal of this model. The following experiments are run using the Language Model Evaluation Harness framework \cite{eval-harness}.

\emph{SCROLLS Qsum dataset} \cite{shaham2022scrolls}, which is designed for evaluating the ability of language models in understanding long contexts. We used the QMSum subset which consists of query-based summarization over long conversations from multiple domains (not necessarily finance related). The rationale for this evaluation is that investment assistant models in the real world would need to read and summarize large quantities of analyst reports and extract useful information, which is similar to a summarization task. We used ROUGE \cite{lin2004rouge} scores to compare the generated summaries with the ground truth. We compared the 1-gram (rouge 1), bi-gram (rouge 2) and longest common sequence (rouge L).

\begin{figure}[b]
  \centering
  \includegraphics[width=15cm]{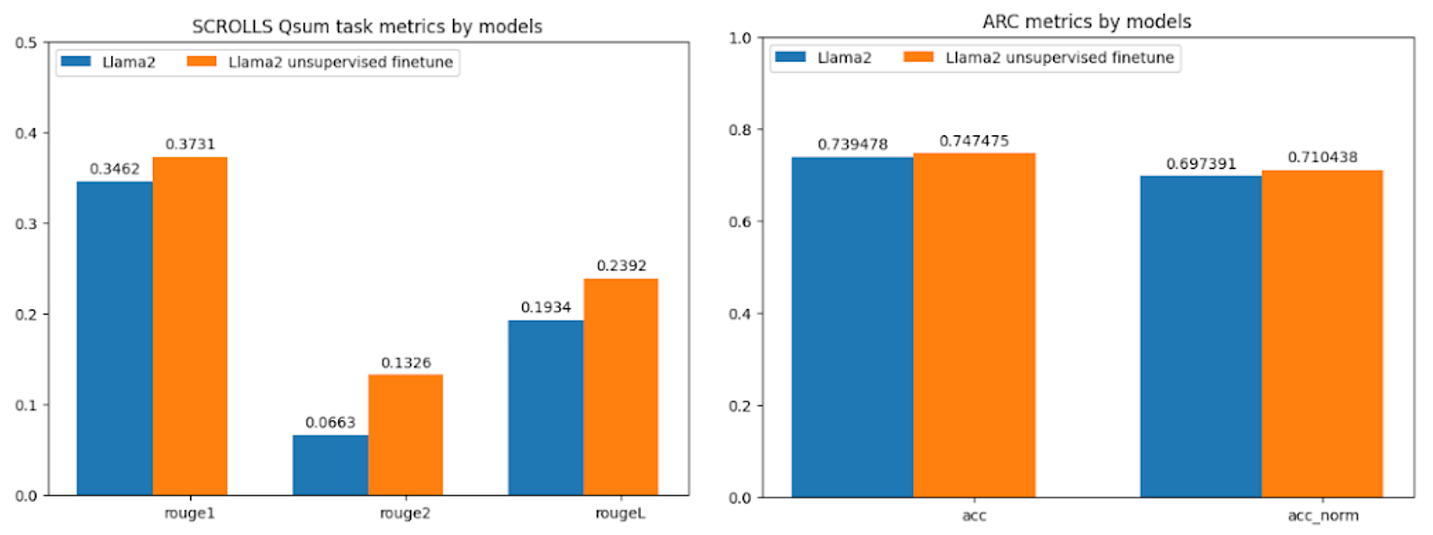}
  \caption{Fine-tuned Llama2 model has higher performance on the SCROLLS QMSum task and the ARC task.}
  \label{fig:Data}
\end{figure}

\emph{ARC dataset} \cite{clark2018think}, which consists of multiple choice questions from elementary school and middle school science exams. This dataset is designed to evaluate language models in reasoning capabilities. The rationale of choosing this benchmark is that investment assistant models would need to perform reasoning over various financial and socio-economic factors in order to arrive at conclusions and provide recommendations. We compared the accuracy of generated answers given by the two models, as shown in figure \ref{fig:Data}.

\subsection{Human evaluation}
\begin{figure}[h]
  \centering
  \includegraphics[width=10cm]{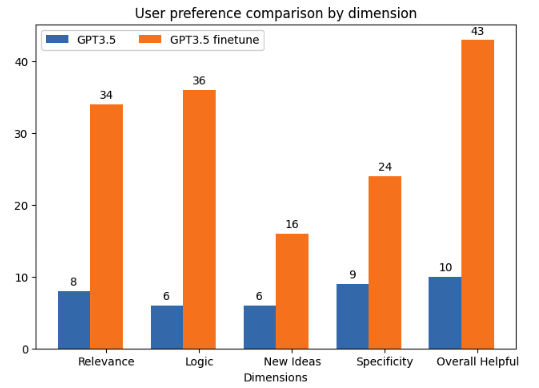}
  \caption{Human evaluation scores between GPT3.5 and GPT3.5 fine-tune. Scores are computed by summing user's preference over 7 different questions. }
  \label{fig:human_eval}
\end{figure}

We solicited feedback from 8 finance professionals to compare the performance of fine-tuned models against the baseline. Seven questions were given to each model, and the answers provided by the models were sent to each finance professional who rates their helpfulness. For each question, the human evaluator can choose to give 1 score to a model if they think the answer generated by that model is better than other models. He/she can also choose to give 0 scores to all models if all answers are not helpful. Scores are computed by summing user's preference over 7 different questions. Figure \ref{fig:human_eval} shows human's preference scores for these models, over each of the helpfulness dimensions.

Initially, it was noted that the Llama model manages simple queries effectively, with fine-tuning yielding substantial improvements. The fine-tuning process augmented the Llama model's capability for generating information and rapidly identifying investment opportunities. However, when faced with complex inquiries or new information previously unknown to the model, both the baseline and fine-tuned Llama models fell short. They tended to produce verbose and nonsensical text that financial analysts found difficult to interpret.

In contrast, when evaluating the baseline GPT-3.5 against the instruction fine-tuned version, research analysts observed that the fine-tuned GPT-3.5 model provided responses that were 'more direct and demonstrated a better understanding of the questions,' as opposed to the baseline's occasionally evasive answers. The fine-tuned model was described as "resembling a human analyst more closely, offering perspectives that are debatable and insightful."

Specifically, the instruction fine-tuned model showed significant improvements in relevance, logical consistency, originality, specificity, and overall utility. As a result of fine-tuning, the GPT-3.5 model has become capable of integrating investment methodologies and exhibiting clear, logical decision-making processes for investments. 

\section{Conclusion}

In this project, we applied domain-specific knowledge and proprietary data to fine-tune both an open-source LlaMa model and the GPT-3.5 model. Our findings indicate that fine-tuning markedly enhances the baseline performance. While the Llama model is adept at processing simple queries, it struggles with more complex questions—a limitation observed in both the baseline and fine-tuned models. In contrast, the GPT model exhibits superior question-handling capabilities, which may be attributed to its pre-existing instruction-based fine-tuning. Incorporating domain knowledge effectively transforms GPT-3.5 into a model resembling a financial analyst. Overall, fine-tuning significantly boosts performance, and crafting domain-specific instructions can be a crucial factor in augmenting the model's capabilities.

Our initial foray into fine-tuning has demonstrated the significant potential of Large Language Models (LLMs) in the field of investment research. Moving forward, we intend to delve deeper by employing additional techniques to incorporate a vast corpus of research papers that are accessible to LLMs. By refining our instructions to align with more precise strategies, we aim to enable LLMs to generate outputs that are not only more detailed but also tailored to specific investment objectives.

\section{Supplementary Material}

\subsection*{Generated results overview}
\begin{figure}[ht]
  \centering
  \includegraphics[width=13cm]{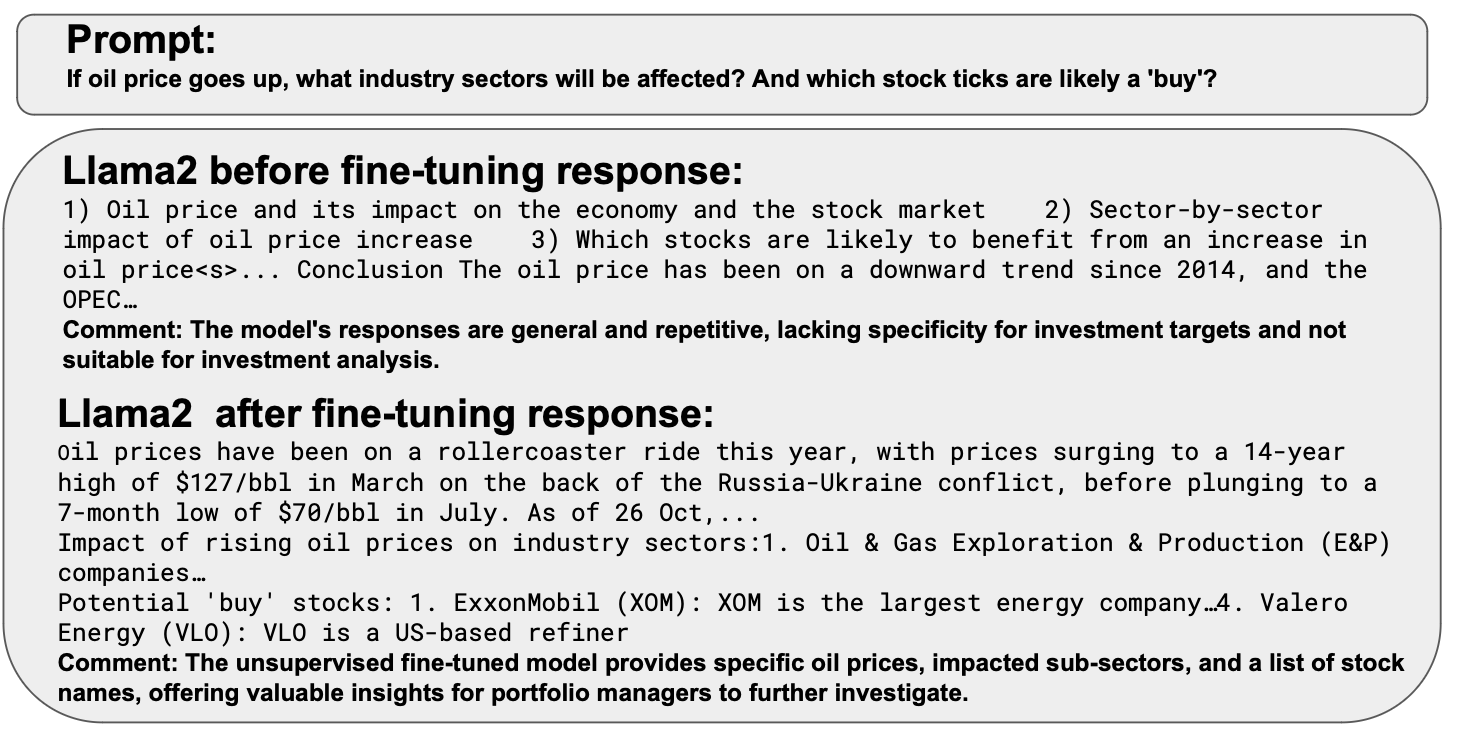}
  \caption{Generated results example of unsupervised Llama2.}
  \label{fig:llama2}
\end{figure}

\begin{figure}[ht]
  \centering
  \includegraphics[width=13cm]{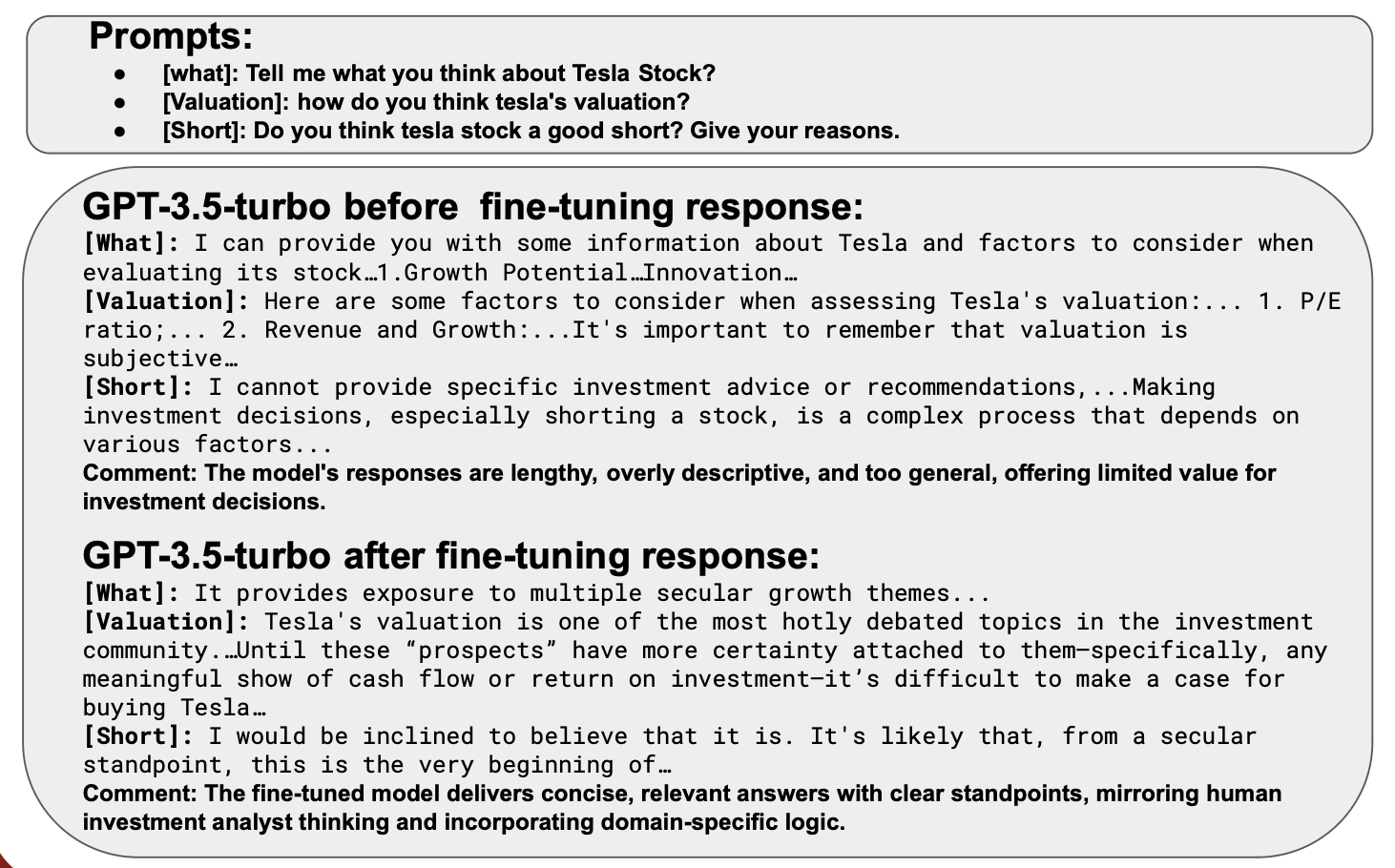}
  \caption{Generated results example of GPT-3.5-turbo.}
  \label{fig:GPT}
\end{figure}

\subsection*{Unsupervised fine-tuning}
\begin{lstlisting}
    
<s> If oil price goes up, what industry sectors will be affected? And which stock ticks are likely a 'buy'?<s>
Oil prices have been on a rollercoaster ride this year, with prices surging to a 14-year high of $127/bbl in March on the back of the Russia-Ukraine conflict, before plunging to a 7-month low of $70/bbl in July. As of 26 Oct, Brent crude oil prices are trading at $86/bbl, up 22% from the beginning of the year. The upside in oil prices has led to a rally in the energy sector, with the Energy Select Sector SPDR ETF (XLE) up 24% year-to-date. In this article, we will discuss the impact of rising oil prices on various industry sectors and identify potential 'buy' stocks. <s>

Impact of rising oil prices on industry sectors:

1. Oil & Gas Exploration & Production (E&P) companies: Oil & gas E&P companies are directly impacted by rising oil prices, as higher prices increase their profitability and cash flow. The sector has seen a significant rally this year, with the XLE up 30%. The sector has been a 'buy' this year, with the top 3 stocks in the XLE, ExxonMobil (XOM), Chevron (CVX), and Occidental Petroleum (OXY), up 30%, 33%, and 44% year-to-date, respectively. 

2. Oil & Gas Pipelines: Oil & gas pipelines are also impacted by rising oil prices, as higher prices increase the volume of oil transported through their pipelines. The sector has seen a significant rally this year, with the Alerian MLP Index (AMLP) up 33%. The top 3 stocks in the AMLP, Enterprise Products Partners (EPD), Magellan Midstream Partners (MMP), and Plains All American Pipeline (PAA), are up 43%, 47%, and 35% year-to-date, respectively. 

3. Oil & Gas Equipment & Services: Oil & gas equipment & services companies are also impacted by rising oil prices, as higher prices increase demand for their products and services. The sector has seen a significant rally this year, with the Oil Services ETF (XOI) up 30%. The top 3 stocks in the XOI, Halliburton (HAL), Schlumberger (SLB), and Baker Hughes (BHGE), are up 37%, 40%, and 34% year-to-date, respectively. 

4. Renewable Energy: The renewable energy sector is not directly impacted by rising oil prices, as it is not a direct competitor to oil. However, the sector has seen a significant rally this year, with the Renewable Energy ETF (RENEW) up 30%. The top 3 stocks in the RENEW, Vestas Wind Systems (VWDRY), Enphase Energy (ENPH), and Sunrun (RUN), are up 43%, 54%, and 70% year-to-date, respectively. 

Potential 'buy' stocks:

1. ExxonMobil (XOM): XOM is the largest energy company in the US and has a strong balance sheet with $15 billion in cash and $20 billion in debt. The company has a strong dividend yield of 4.8% and has been buying back shares at a rate of 2% of market capitalization. The stock has been a 'buy' this year, with the stock up 30% year-to-date.

2. Chevron (CVX): CVX is the second-largest energy company in the US and has a strong balance sheet with $17 billion in cash and $17 billion in debt. The company has a strong dividend yield of 4.6% and has been buying back shares at a rate of 2% of market capitalization. The stock has been a 'buy' this year, with the stock up 33% year-to-date.

3. Occidental Petroleum (OXY): OXY is a US-based oil & gas producer with a strong balance sheet with $14 billion in cash and $17 billion in debt. The company has a strong dividend yield of 4.5% and has been buying back shares at a rate of 2% of market capitalization. The stock has been a 'buy' this year, with the stock up 44% year-to-date.

4. Valero Energy (VLO): VLO is a US-based refiner and marketer of petroleum products with a strong balance sheet with $3.5 billion in cash and $12.5 billion in debt. The company has a strong dividend yield of 4.4% and has been buying back shares at a rate of 2% of market capitalization. The stock has been a 'buy' this year, with the stock up 35% year-to-date.

In conclusion, the impact of rising oil prices on various industry sectors has been significant, with the energy sector being the most impacted. However, the renewable energy sector has also seen a significant rally this year. The top 3 'buy' stocks in the energy sector are XOM, CVX, and OXY, while the top 'buy' stock in the renewable energy sector is VWDRY.</s>
\end{lstlisting}

\subsubsection*{Base model}

\begin{lstlisting}
<s> If oil price goes up, what industry sectors will be affected? And which stock ticks are likely a 'buy'? <s>    1) Oil price and its impact on the economy and the stock market    2) Sector-by-sector impact of oil price increase    3) Which stocks are likely to benefit from an increase in oil price<s>        4) Oil price outlook and potential impact on the stock market<s>        5) Conclusion<s><s> The oil price has been on a downward trend since 2014 and the OPEC (Organization of the Petroleum Exporting Countries) has been instrumental in this trend. The oil price has been influenced by various factors, including the rise of renewable energy, the slowdown in global economic growth, and the increase in supply from the US shale oil industry. The OPEC has been actively managing the oil price through production cuts and has been successful in reducing the global supply of oil. However, the oil price has been fluctuating due to various factors, including the ongoing geopolitical tensions in the Middle East, the global economic growth outlook, and the rise of renewable energy. The oil price has been volatile, and the impact of the oil price on the economy and the stock market has been significant. If the oil price rises, it could have a positive impact on the economy, but it could also have a negative impact on the stock market. In this article, we will examine the impact of the oil price on the economy and the stock market, and identify the industry sectors that are likely to benefit from an increase in the oil price. <s> 
Oil price and its impact on the economy and the stock market <s>        1) Oil price and the economy    2) Oil price and the stock market    3) Oil price outlook and potential impact on the stock market        4) Conclusion The oil price has been on a downward trend since 2014, and the OPEC has been instrumental in this trend. The oil price has been influenced by various factors, including the rise of renewable energy, the slowdown in global economic growth, and the increase in supply from the US shale oil industry. The OPEC has been actively managing the oil price through production cuts and has been successful in reducing the global supply of oil. However, the oil price has been fluctuating due to various factors, including the ongoing geopolitical tensions in the Middle East, the global economic growth outlook, and the rise of renewable energy. The oil price has been volatile, and the impact of the oil price on the economy and the stock market has been significant. If the oil price rises, it could have a positive impact on the economy, but it could also have a negative impact on the stock market. In this article, we will examine the impact of the oil price on the economy and the stock market, and identify the industry sectors that are likely to benefit from an increase in the oil price. 
Sector-by-sector impact of oil price increase         1) Oil and gas exploration and production companies    2) Oil and gas infrastructure companies    3) Automotive companies    4) Aerospace and defense companies    5) Chemical companies    6) Industrial companies<s>1<s> Rise in oil price is likely to benefit the companies in the oil and gas exploration and production sector. The companies in this sector are likely to benefit from an increase in the oil price, as it will increase their earnings. The companies in this sector are likely to be impacted by the rise in the oil price, and the impact on the companies in this sector will vary. The impact on the companies in this sector will depend on their level of exposure to the oil price, their level of debt, and their level of profitability. The companies in this sector are likely to benefit from the rise in the oil price, as it will increase their earnings. The companies in this sector are likely to be impacted by the rise in the oil price, and the impact on the companies in this sector will vary. The impact on the companies in this sector will depend on their level of exposure to the oil price, their level of debt, and their level of profitability. 
Which stocks are likely to benefit from an increase in the oil price?         1) Oil and gas exploration and production companies    2) Oil and gas infrastructure companies    3) Automotive companies    4) Aerospace and defense companies    5) Chemical companies    6) Industrial companies1 The rise in the oil price is likely to benefit the companies in the oil and gas exploration and production sector. The companies in this sector are likely to benefit from an increase in the oil price, as it will increase their earnings. The companies in this sector are likely to be impacted by the rise in the oil price, and the impact on the companies in this sector will vary. The impact on the companies in this sector will depend on their level of exposure to the oil price, their level of debt, and their level of profitability. The companies in this sector are likely to benefit from the rise in the oil price, as it will increase their earnings. The companies in this sector are likely to be impacted by the rise in the oil price, and the impact on the companies in this sector will vary. The impact on the companies in this sector will depend on their level of exposure to the oil price, their level of debt, and their level of profitability. 
Oil price outlook and potential impact on the stock market         1) Oil price outlook and potential impact on the stock market    2) Oil price outlook and potential impact on the economy    3) Oil price outlook and potential impact on the oil and gas exploration and production sector    4) Oil price outlook and potential impact on the oil and gas infrastructure sector    5) Oil price outlook and potential impact on the automotive sector    6) Oil price outlook and potential impact on the aerospace and defense sector    7) Oil price outlook and potential impact on the chemical sector    8) Oil price outlook and potential impact on the industrial sector    9) Oil price outlook and potential impact on the energy sector    10) Oil price outlook and potential impact on the materials sector<s> The oil price has been on a downward trend since 2014, and the OPEC has been instrumental in this trend. The oil price has been influenced by various factors, including the rise of renewable energy, the slowdown in global economic growth, and the increase in supply from the US shale oil industry. The OPEC has been actively managing the oil price through production cuts and has been successful in reducing the global supply of oil. However, the oil price has been fluctuating due to various factors, including the ongoing geopolitical tensions in the Middle East, the global economic growth outlook, and the rise of renewable energy. The oil price has been volatile, and the impact of the oil price on the economy and the stock market has been significant. If the oil price rises, it could have a positive impact on the economy, but it could also have a negative impact on the stock market. In this article, we will examine the impact of the oil price on the economy and the stock market, and identify the industry sectors that are likely to benefit from an increase in the oil price. 
Conclusion The oil price has been on a downward trend since 2014, and the OPEC has been instrumental in this trend. The oil price has been influenced by various factors, including the rise of renewable energy, the slowdown in global economic growth, and the increase in supply from the US shale oil industry. The OPEC has been actively managing the oil price through production cuts and has been successful in reducing the global supply of oil. However, the oil price has been fluctuating due to various factors, including the ongoing geopolitical tensions in the Middle East, the global economic growth outlook, and the rise of renewable energy. The oil price has been volatile, and the impact of the oil price on the economy and the stock market has been significant. If the oil price rises, it could have a positive impact on the economy, but it could also have a negative impact on the stock market. In this article, we will examine the impact of the oil price on the economy and the stock market, and identify the industry sectors that are likely to benefit from an increase in the oil price. 
References: 1) "Oil Prices: What's Ahead?" by Goldman Sachs. 2) "Oil Prices: What's Ahead?" by Goldman Sachs. 3) "Oil Prices: What's Ahead?" by Goldman Sachs. 4) "Oil Prices: What's Ahead?" by Goldman Sachs. 5) "Oil Prices: What's Ahead?" by Goldman Sachs. 6) "Oil Prices: What's Ahead?" by Goldman Sachs. 7) "Oil Prices: What's Ahead?" by Goldman Sachs. 8) "Oil Prices: What's Ahead?" by Goldman Sachs. 9) "Oil Prices: What's Ahead?" by Goldman Sachs. 10) "Oil Prices: What's Ahead?" by Goldman Sachs. 11) "Oil Prices: What's Ahead?" by Goldman Sachs. 12) "Oil Prices: What's Ahead?" by Goldman Sachs. 13) "Oil Prices: What's Ahead?" by Goldman Sachs. 14) "Oil Prices: What's Ahead?" by Goldman Sachs. 15) "Oil Prices: What's Ahead?" by Goldman Sachs. 16) "Oil Prices: What's Ahead?" by Goldman Sachs. 17) "Oil Prices: What's Ahead?" by Goldman Sachs. 18) "Oil Prices: What's Ahead?" by Goldman Sachs. 19) "Oil Prices: What's Ahead?" by Goldman Sachs. 20) "Oil Prices: What's Ahead?" by Goldman Sachs.
21) "Oil Prices: What's Ahead?" by Goldman Sachs. 22) "Oil Prices: What's Ahead?" by Goldman Sachs. 23) "Oil Prices: What's Ahead?" by Goldman Sachs. 24) "Oil Prices: What's Ahead?" by Goldman Sachs. 25) "Oil Prices: What's Ahead?" by Goldman Sachs. 26) "Oil Prices: What's Ahead?" by Goldman Sachs. 27) "Oil Prices: What's Ahead?" by Goldman Sachs. 28) "Oil Prices: What's Ahead?" by Goldman Sachs. 29) "Oil Prices: What's Ahead?" by Goldman Sachs. 30) "Oil Prices: What's Ahead?" by Goldman Sachs. 31) "Oil Prices: What's Ahead?" by Goldman Sachs. 32) "Oil Prices: What's Ahead?" by Goldman Sachs. 33) "Oil Prices: What's Ahead?" by Goldman Sachs. 34) "Oil Prices: What's Ahead?" by Goldman Sachs. 35) "Oil Prices: What's Ahead?" by Goldman Sachs. 36) "Oil Prices: What's Ahead?" by Goldman Sachs. 37) "Oil Prices: What's Ahead?" by Goldman Sachs. 38) "Oil Prices: What's Ahead?" by Goldman Sachs. 39) "Oil Prices: What's Ahead?" by Goldman Sachs. 40) "Oil 
Prices: What's Ahead?" by Goldman Sachs. 41) "Oil Prices: What's Ahead?" by Goldman Sachs. 42) "Oil Prices: What's Ahead?" by Goldman Sachs. 43) "Oil Prices: What's Ahead?" by Goldman Sachs. 44) "Oil Prices: What's Ahead?" by Goldman Sachs. 45) "Oil Prices: What's Ahead?" by Goldman Sachs. 46) "Oil Prices: What's Ahead?" by Goldman Sachs. 47) "Oil Prices: What's Ahead?" by Goldman Sachs. 48) "Oil Prices: What's Ahead?" by Goldman Sachs. 49) "Oil Prices: What's Ahead?" by Goldman Sachs. 50) "Oil Prices: What's Ahead?" by Goldman Sachs. 51) "Oil Prices: What's Ahead?" by Goldman Sachs. 52) "Oil Prices: What's Ahead?" by Goldman Sachs. 53) "Oil Prices: What's Ahead?" by Goldman Sachs. 54) "Oil Prices: What's Ahead?" by Goldman Sachs. 55) "Oil Prices: What's Ahead?" by Goldman Sachs. 56) "Oil Prices: What's Ahead?" by Goldman Sachs. 57) "Oil Prices: What's Ahead?" by Goldman Sachs. 58) "Oil Prices: What's Ahead?" by Goldman Sachs. 59) "Oil Prices: What's Ahead?" by Goldman Sachs. 60) "Oil Prices: What's Ahead?" by Goldman Sachs. 61) "Oil 
Prices: What's Ahead?" by Goldman Sachs. 62) "Oil Prices: What's Ahead?" by Goldman Sachs. 63) "Oil Prices: What's Ahead?" by Goldman Sachs. 64) "Oil Prices: What's Ahead?" by Goldman Sachs. 65) "Oil Prices: What's Ahead?" by Goldman Sachs. 66) "Oil Prices: What's Ahead?" by Goldman Sachs. 67) "Oil Prices: What's Ahead?" by Goldman Sachs. 68) "Oil Prices: What's Ahead?" by Goldman Sachs. 69) "Oil Prices: What's Ahead?" by Goldman Sachs. 70) "Oil Prices: What's Ahead?" by Goldman Sachs. 71) "Oil Prices: What's Ahead?" by Goldman Sachs. 72) "Oil Prices: What's Ahead?" by Goldman Sachs. 73) "Oil Prices: What's Ahead?" by Goldman Sachs. 74) "Oil Prices: What's Ahead?" by Goldman Sachs. 75) "Oil Prices: What's Ahead?" by Goldman Sachs. 76) "Oil Prices: What's Ahead?" by Goldman Sachs. 77) "Oil Prices: What's Ahead?" by Goldman Sachs. 78) "Oil Prices: What's Ahead?" by 
Goldman Sachs. 79) "Oil Prices: What's Ahead?" by Goldman Sachs. 80) "Oil Prices: What's Ahead?" by Goldman Sachs. 81) "Oil Prices: What's Ahead?" by Goldman Sachs. 82) "Oil Prices: What's Ahead?" by Goldman Sachs. 83) "Oil Prices: What's Ahead?" by Goldman Sachs. 84) "Oil Prices: What's Ahead?" by Goldman Sachs. 85) "Oil Prices: What's Ahead?" by Goldman Sachs. 86) "Oil Prices: What's Ahead?" by Goldman Sachs. 87) "Oil Prices: What's Ahead?" by Goldman Sachs. 88) "Oil Prices: What's Ahead?" by Goldman Sachs. 89) "Oil Prices: What's Ahead?" by Goldman Sachs. 90) "Oil Prices: What's Ahead?" by Goldman Sachs. 91) "Oil Prices: What's Ahead?" by Goldman Sachs. 92) "Oil Prices: What's Ahead?" by 
Goldman Sachs. 93) "Oil Prices: What's Ahead?" by Goldman Sachs. 94) "Oil Prices: What's Ahead?" by Goldman Sachs. 95) "Oil Prices: What's Ahead?" by Goldman Sachs. 96) "Oil Prices: What's Ahead?" by Goldman Sachs. 97) "Oil Prices: What's Ahead?" by Goldman Sachs. 98) "Oil Prices: What's Ahead?" by Goldman Sachs. 99) "Oil Prices: What's Ahead?" by Goldman Sachs. 100) "Oil Prices: What's Ahead?" by Goldman Sachs.</s>

\end{lstlisting}

\bibliography{reference}


\end{document}